\documentclass[aps,prl,twocolumn,showpacs]{revtex4}

\usepackage{graphicx,color}% Include figure files
\usepackage{soul}

\def\be{\begin{equation}}
\def\ee{\end{equation}}
\def\bc{\begin{center}}
\def\ec{\end{center}}

\begin{document}

\title{Noise induced rupture process: Phase boundary and scaling 
of waiting time distribution}
\author{Srutarshi Pradhan}
\email{srutarshi.pradhan@sintef.no;srutarshi@gmail.com }
\affiliation{Sintef Petroleum Research, N-7465 Trondheim, Norway}
\author{Anjan Kumar Chandra}
\email{anjanphys@gmail.com}
\author{Bikas K. Chakrabarti}
\email{bikask.chakrabarti@saha.ac.in}
\affiliation{Saha Institute of Nuclear Physics, 1/AF Bidhannagar, Kolkata, India}

\begin{abstract}
A bundle of fibers has been considered here as a model for composite materials,
 where breaking of the fibers occur due to a combined influence of applied load
 (stress) and external noise. 
Through numerical simulation and a mean-field calculation we show 
that there exists a robust phase boundary between continuous 
(no waiting time) and intermittent fracturing regimes. 
In the intermittent regime, 
 throughout the entire rupture process avalanches of 
different sizes are produced  and there is a waiting time between two 
consecutive avalanches. 
The statistics of waiting times follows a Gamma distribution and the avalanche 
distribution shows power law scaling, similar to what have 
been observed in case of earthquake events and bursts in fracture experiments.
We propose a prediction scheme that can tell when the system is expected to 
reach the continuous fracturing point from the intermittent phase.   
\end{abstract}
\pacs{02.50.-r, 05.40.-a, 91.30.-f}
\maketitle

Rupture and breakdown \cite{CB97,HR90} are complex processes that 
occur both in micro and macro scales. 
Natural rupture phenomena like earthquake, land-slide, mine-collapse, 
snow-avalanches often appear catastrophic to human society. It is 
therefore a 
fundamental challenge to understand the underlying rupture process 
so that the losses in terms of properties and lives can be 
minimised by providing early alarms. 
The same crisis persists in construction engineering and material 
industry where detail knowledge of the strength of the materials and their 
failure properties, are essential. But the physical processes which initiate 
rupture, help its growth 
and finally results in breakdown, are not completely understood yet.
     
Fiber bundle model (FBM) has become a useful tool for studying rupture 
and failure \cite{RMP} 
of  composite materials under different loading conditions. The simple geometry 
of the model and clear-cut load-sharing rules allow to achieve analytic 
solutions \cite{PC01,PBC02,BPC03}
 to an  extent that is not possible in any of the fracture models studied 
so far by the fracture community. 
FBM was introduced first in connection with textile 
engineering \cite{Peirce} and recently physicists took interest in it, mainly 
to explore the 
critical failure dynamics and avalanche phenomena in this model
 \cite{HH,PHH,PH}. Not only
the classical fracture-failure (stress-induced) in composites, FBM has been 
used successfully for studying noise-induced (fatigue) failure 
\cite{Lawn,Coleman,Ciliberto,Roux,Pradhan}, creep \cite{HKH02,nhgs05,BK08} and thermally induced failures \cite{YKI10,YKI12}. 
The statistics of avalanches in these type of failure models show 
similarities with results for acoustic emissions \cite{AE} (during material 
failure) and 
earthquakes \cite{Corral06,Corral03, Bak}.    

In this work, through  waiting time and avalanche statistics, we analyze
 a noise induced intermittent fracturing process in composite materials under 
fixed external loading.
The waiting time is defined as the time (Monte-Carlo steps) between two 
consecutive avalanches 
 in the avalanche time series for the entire failure process.
Through a mean-field calculation we show that
in the stress-noise space, there exists a robust 
phase boundary between continuous 
(no waiting time) and intermittent fracturing regimes and that  can be 
verified by numerical simulations. 
In the intermittent fracturing regime we study the distributions of avalanches 
and  waiting times for different type of fiber strength distributions. Finally
we mention and discuss studies on waiting-time statistics  in other fracture 
models, earthquake events and fracture experiments.
  
We consider first a bundle of $N$ parallel fibers -
and a load ($W=\sigma N$) is applied on the bundle. The fibers have different
individual strengths ($x$) which are drawn from a probability distribution and 
the bundle has a critical strength $\sigma_c$ \cite{RMP}, so that
without any noise, the bundle 
does not fail completely for stress $\sigma \le \sigma_c$, but it fails 
immediately for $\sigma>\sigma_c$. We now assume that each fiber having 
strength $x_i$ has
a finite probability $P(\sigma,T)$ of failure at any stress
$\sigma$ induced by a nonzero noise $T$:
\begin{equation}
\label{july31-1}
P(\sigma ,T)=\left\{ \begin{array}{cc}
\exp \left[ -\frac{1}{T}\left( \frac{x_i}{\sigma }-1\right) \right] , & 0\leq \sigma \leq x_i\\
1, & \sigma >x_i
\end{array}\right .
\end{equation}
Here $P(\sigma,T)$ increases as $T$ increases and for a fixed value of 
$T$ and $\sigma_c$, as we increase $\sigma$, the bundle breaks more rapidly. 
We simulate this failure phenomenon following Eq.~(\ref{july31-1}) in discrete 
time $t$. 
After each failure (at the fixed stress $\sigma$), the total load $N\sigma$
is redistributed among the remaining fibers equally and we check at time
$t+1$, if the present stress $\sigma(t+1) = W/N(t+1)$ can induce any further 
failure
following Eq.~(\ref{july31-1}). When the value of $\sigma$ is considerably
large, it so happens that at every time step at least a 
single fiber breaks until the complete collapse of the bundle. This is
a single avalanche and there is no waiting time \cite{Pradhan}. 
But as we decrease
the initial value of $\sigma$, at a limiting value, in a particular 
time step $t$ not a single fiber breaks. We consider this as a single waiting 
time ($t_W = 1$) and the limiting value of $\sigma$, at which the
waiting time appears for the first time is denoted by $\sigma_0$. This is 
the onset of intermittent fracturing process. 
After one waiting time, again another avalanche starts and 
eventually all the fibers break after such finite number of avalanches.
The number of fibers broken during a single avalanche is counted as the
avalanche size ($m$). 
It is obvious that as we increase the value 
of $T$, the value of $\sigma_0$ decreases. When the noise is large,
the initial applied load has to be smaller for the emergence of a waiting time. 
Thus stress ($\sigma$) and noise ($T$) values determine whether the system is 
in continuous rupture phase or in the intermittent rupture phase. It may be 
mentioned that  $T$ can be interpreted  as a measure of thermal noise in the 
system and similar thermally activated breakdown in fiber bundle model had been 
studied experimentally \cite{Ciliberto} and theoretically \cite{Roux}. 
\begin{figure}
\includegraphics{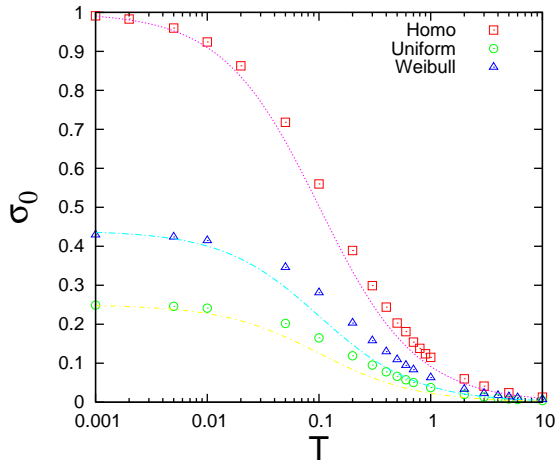}
\caption{\label{fig:fg1}
Phase boundary ($\sigma_0$ vs. $T$ plot) for three different type of fiber 
strength distributions with $N=20000$. Data points are simulation results (averages are taken over $100$ samples) and 
solid lines are analytic estimates (Eqs. 3,4) based on mean-field arguments.}
\end{figure}

To determine the phase boundary we can give a mean-field argument that at 
$\sigma = \sigma_0$, at least one 
fiber must break to trigger the continuous fracturing process. After this 
single failure the load has to be redistributed on the intact fibers and the 
effective stress must be more than $\sigma_0$ - which in turn enhances 
failure probability for all the intact fibers. 
Therefore in case of homogeneous bundle where all the fibers have identical 
strength $x_i=1$ (therefore $\sigma_c=1$), at the phase boundary
$N P(\sigma_0,T) \ge 1$
giving
\be
N \exp\left[-\frac{1}{T}\left(\frac{1}{\sigma_0}-1\right)\right] \ge 1
\ee
which gives
\be 
\sigma_0 \ge \frac{1}{1-T log(1/N)}.
\label{sigma0}
\ee
In the absence of noise $T$, $\sigma_0=1=\sigma_c$, which is consistent with the 
static FBM results \cite{RMP}.  
This analytic estimate coincides with the data obtained from simulation 
(Fig.~\ref{fig:fg1}). It shows a nice phase boundary between the continuous 
and intermittent fracturing regimes. 

For heterogeneous cases where fibers have different 
strength and the 
whole bundle has a critical strength $\sigma_c$, we make the conjecture that 
\be 
\sigma_0 \ge \frac{\sigma_c}{1-T log(1/N)};
\ee
keeping in mind that in absence of noise $T$, $\sigma_0=\sigma_c$.
To verify our conjecture we choose heterogeneous bundles of $N$ fibers
 where strength of the fibers are drawn from a statistical distribution.
We have considered two different kinds of fiber strength distributions: 
(1) uniform distribution of fiber strength having cumulative form 
$Q(x)=x$ for $0<x\le1$
and (2) Weibull distribution  
$Q(x)=1- exp (-x^k)$
where $k$ is the Weibull index 
(we have taken $k = 2.0$ and $5.0$). Each fiber has a finite probability 
$P(\sigma,T)$ of failure at any stress
$\sigma$ induced by a nonzero $T$ as mentioned before. Similar to
the homogeneous case, for a particular value of $T$, below a certain
value of $\sigma$, the waiting time appears here. 
One can see that the theoretical estimate of phase boundary agree with 
the numerical data for the heterogeneous cases (Fig. \ref{fig:fg1}). 
However this agreement was much better for homogeneous case. This difference
 can be 
explained through the amount of randomness involved in the respective systems. 
In case 
of homogeneous bundle there is no randomness in the fiber strength - the only 
randomness is coming from the noise term.  Whereas in case of heterogeneous 
bundles - there are two sources of randomness - in the fiber strengths and 
in the noise term. 

\begin{figure}
\includegraphics{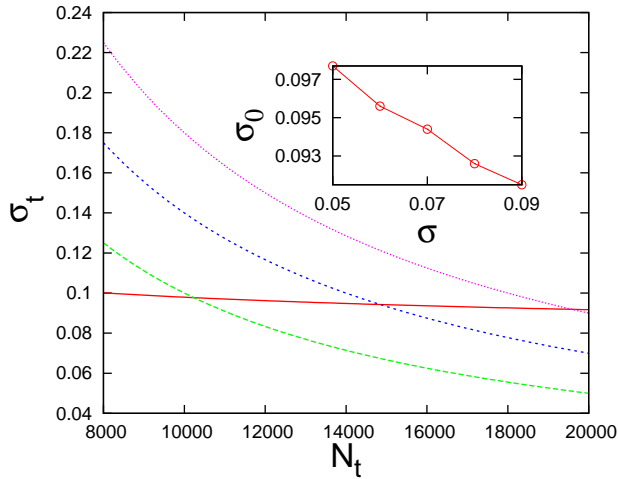}
\caption{\label{prediction}
Prediction scheme for homogeneous bundle: $\sigma_t$ vs. $N_t$ are plotted for 
three different initial stress values  with $N=20000$. The intersection points 
with red line (equation \ref{sigma0}) indicates the starting of continuous 
fracturing. The inset shows the variation of $\sigma_0$ value with initial 
stress ($\sigma$) in the same bundle.}
\end{figure}

The mean-field calculation (\ref{sigma0}) suggests that $\sigma_0$ value 
depends on the number of intact fibers in the bundle ($N$). It 
increases with decrease in the number of intact fibers at time $t$ ($N_t$).
Therefore, when we 
start with a much lower stress value 
($\sigma<\sigma_0$) the $\sigma_0$ value 
  increases slightly with time (as $N_t$ decreases with small individual
failures). But the effective stress value follows a strict relation with 
applied stress and number of intact fibers as       
\be 
\sigma_t = \sigma N/N_t.
\label{sigmat}
\ee
These two equations (\ref{sigma0} and \ref{sigmat}) allow us to 
 make a 
theoretical prediction of $\sigma_0$ value for a 
particular bundle of 
homogeneous fibers. If we plot together -$\sigma_0$ vs. $N_t$ and $\sigma_t$ 
vs. $N_t$ for a particular $\sigma$  - then the point of intersection will 
give the $\sigma_0$ value for that particular $\sigma$ value 
(Fig. \ref{prediction}). Therefore, during a fracturing process if we can 
measure the effective stress or the number of intact elements in the system, 
we can always predict the onset of continuous fracturing.   

Existence of such a phase boundary has important consequences on 
fracturing study in material failure and other fracture-breakdown phenomena.   
In real situations of material/rock fracturing, acoustic emission measurements
can show clearly whether an ongoing fracturing process belongs to continuous or 
intermittent fracturing phase. Acoustic emissions \cite{AE} are basically  
sound  waves produced during micro-crack opening within the material body due 
to external stress and noise factors. Once a system enters into continuous 
fracturing phase the breakdown must be imminent. Thus the identification of
rupture phase can predict the fate of a system correctly.        

\begin{figure}
\noindent \includegraphics{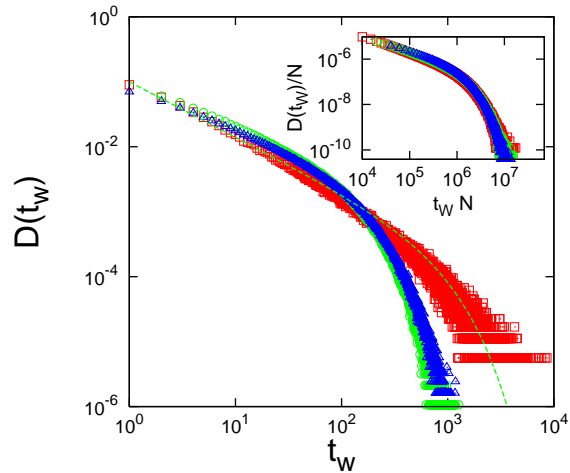}
\caption{The simulation results for the waiting time distributions for 
 three different type of fiber strength distributions (square, circle and 
triangle symbols are used for homogeneous, uniform and Weibull distributions 
respectively) with 
$N=20000$. All the curves can be fitted  with the Gamma form $exp(-t_W/a)/t_W^{1-\gamma}$ (dashed line)
where  $\gamma = 0.15$ for homogeneous case (averages are taken over $25$ samples) and  $\gamma = 0.26$ for 
uniform and Weibull distributions (averages are taken over $100$ samples). 
In the inset we show the data collapse of the waiting time distributions with 
system sizes for uniform distribution.} 
\label{fig:fg2}
\end{figure}

In the intermittent fracturing phase avalanches of different sizes are produced
separated by waiting times ($t_W$) of different magnitudes.
This happens for a stress value $\sigma$ below $\sigma_0$ at a certain noise 
($T$) level.
We have studied the waiting time 
distribution for both homogeneous and heterogeneous bundles with $N = 20000$.
Each curve can be fitted with a Gamma distribution \cite{Corral06,Corral03,Bak}
\be 
D(t_W) \propto \exp(-t_W/a)/t_W^{1-\gamma}
\label{gamma_dist}
\ee 
where $\gamma = 0.15$ for homogeneous case and  $\gamma = 0.26$ for 
heterogeneous cases (Fig.~\ref{fig:fg2}). 
As shown in the inset of Fig.~\ref{fig:fg2}, the plot of $D(t_W)/N$ 
against $t_W N$ gives good data collapse for different $N$ values 
($D(t_W)=A(1-P)^{t_WN}\sim A exp(-Pt_WN),$
where $P$ denotes individual failure probability and  $A$ is a constant, hence 
the normalization of $D(t_W)$ requires $A\sim N$).
Such a data 
collapse indicates the robustness of the Gamma function form.
The value of $a$ is the measure of the extent of the power law
regime and it has different values for different types of strength 
distribution.
As we increase $N$, the value of $a$ gradually decreases. 
We have also studied the waiting time distribution for a fixed value of $N$, 
but different sets of values of $T$ and $\sigma$,
all of which shows Gamma  distribution of the form of Eq.~\ref{gamma_dist}.
For a fixed value of $N$ and $T$ as $\sigma$ decreases, the power law region 
extends longer
and thus the value of $a$ increases, but the exponent of power law decay 
remains same. Again for a certain value of
$N$ and $\sigma$ as $T$ decreases, the value of $a$ increases without any
change in the power law exponent. 
These results imply that the 
power law exponent  remains unchanged with variation of $\sigma$, $T$ and $N$. 

The noise-induced rupture process, modeled here, has two basic ingredients, 
external stress $\sigma$ and noise $T$. The noise term triggers initial 
rupture which induces one or more load-redistribution cycles that finally 
enhances the effective stress level on the system. Therefore the initial 
phase 
of the rupture process is dominated by noise term and as the rupture process
goes on stress factor becomes more dominating.  At the final stage the stress 
redistribution mechanism drives the system toward complete collapse through a 
big avalanche. 

\begin{figure}[h]
\noindent \includegraphics {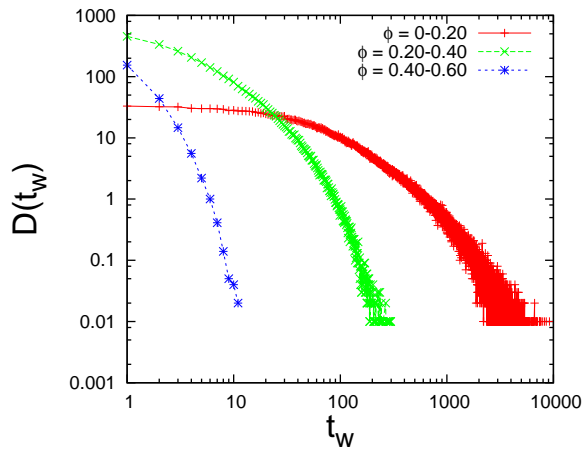}
\caption{ Evolution of waiting time for a homogeneous fiber bundle ($N = 20000$, average number = $25$) at $T = 0.9$ and $\sigma=0.062$.}
\label{fig:homogen.evol}
\end{figure}

\begin{figure}[h]
\noindent \includegraphics {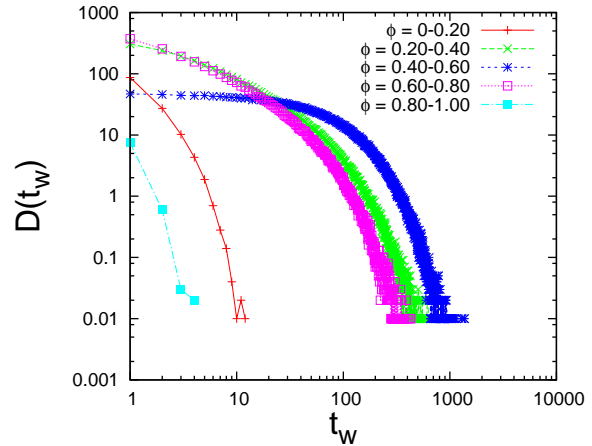}
\caption{Evolution of waiting time for  uniform fiber strength distribution
 ($N = 20000$, average number = $100$) at $T=0.7$ and $\sigma=0.027$.}
\label{fig:uniform.evol}
\end{figure}

For finite values of $N$,
we have studied the waiting time distribution at an interval of $0.20$ of 
the fraction of the broken fibers ($\phi$). It has been observed that within 
the intermittent
regime for homogeneous fiber bundle ($N=20000$) the waiting time distribution is purely
a Gamma distribution during the first $0.20$ fraction of fibers broken
(Fig.~\ref{fig:homogen.evol}).
During the next $0.20$ fraction of broken fibers (i.e., 0.20-0.40), the
power law portion
diminishes and for the next interval (0.40-0.60) there is no power law
regime at all.
For the next two intervals ($0.60-0.80$ and $0.80-1.00$) no waiting time
appears
which implies that for homogeneous fiber bundle the waiting time monotonously
disappears with the breaking of fibers (Fig.~\ref{fig:homogen.evol}).
The nature of evolution of waiting time distribution for the uniform
distribution is different from that of the homogeneous one. In case of
uniform fiber bundle upto $0.60$ fraction of fibers (at an interval of
$0.20$) the value of $a$ increases and large waiting times appear as
more fibers break. This is due to the fact that initially the fibers of very
low strength breaks down instantaneously as soon as a finite stress is
applied. But gradually those fibers of low strength become scarce and due to
the presence of fibers of intermediate strengths and the moderately increased
stress (due to gradual breakdown of fibers), waiting times of broad range
appear. But the breaking of the consequent fibers are faster
due to the increased stress and gradually the $a$ value decreases (Fig.~\ref{fig:uniform.evol}).

\begin{figure}
\includegraphics{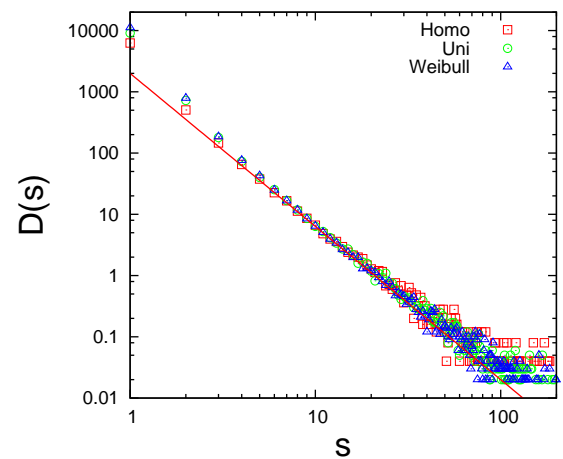}
\caption{\label{fig:avsize}
 Numerical data for avalanche size distributions for three different type of 
fiber threshold distributions (averages are taken over $25$ samples for 
homogeneous case and $100$ samples for uniform and Weibull cases) with 
$N=20000$. The straight line has a slope $-2.5$.}
\end{figure}

In general, avalanches or bursts  bear important information of the dynamics of 
intermittent processes. In our model the noise $T$ triggers a rupture 
process which continues through load (or stress) redistribution mechanism.
The avalanche size distributions follow an universal power law 
($D(s)\sim s^{-\xi}$) scaling with 
exponent $\xi =2.5$. This result (Fig.\ref{fig:avsize}) demands that such 
intermittent rupture process 
belong to the quasi-static fracturing class, where the universality of the 
exponent value has 
already been established \cite{HH}.  

\begin{figure}
\noindent \includegraphics {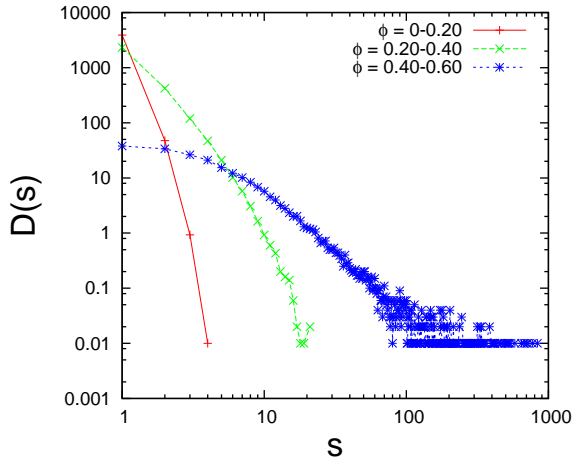}
\caption{ Evolution of the avalanche distributions with the fraction of broken 
fibers ($\phi$) for a homogeneous fiber bundle ($N = 20000$, averages are taken over $25$ samples) at $T = 0.9$ and $\sigma=0.062$.}
\label{fig:avhomo-evolv}
\end{figure}

\begin{figure}
\noindent \includegraphics {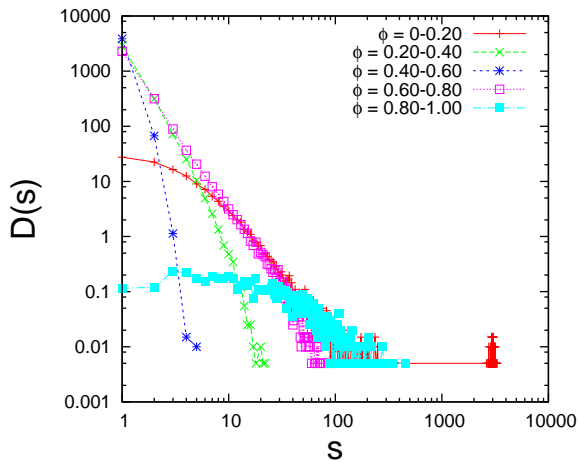}
\caption{Evolution of avalanche distributions with fraction of broken fibers
($\phi$) for  uniform fiber strength distribution
 ($N = 20000$, averages are taken over $100$ samples) at $T=0.7$ and 
$\sigma=0.027$.}
\label{fig:avuni-evolv}
\end{figure}
Instead of considering all the avalanches up to the complete failure of the 
system, if we gather avalanches within some window during the breaking process, 
the shape of the avalanche distributions  change as the system approaches 
complete failure. In case of homogenious strength distribution, there is a 
monotonic variation (Fig. \ref{fig:avhomo-evolv}), i.e., more and more large 
avalanches appear as the failure point is approached but bundle with  uniform 
strength distribution shows a non-monotonic variation (Fig. \ref{fig:avuni-evolv}).        
  
Our model for noise induced rupture process is not limited to any particular 
system, rather it is a general approach and can model more complex situations.
 There are evidences of stress redistribution 
and stress-localisation around fracture/fault lines 
and several factors that can help rupture evolution are friction, plasticity, 
fluid migration, spatial heterogeneities, chemical reactions etc. In our model
  such stress redistribution/localisation can be taken into account through a 
proper load sharing scheme and noise term ($T$) can represent the combined 
effect of other factors.    

We would like to mention here that waiting times and their statistics in different types of fracture models have also been discussed recently.    
Creep rupture in a non-linear viscoelastic FBM  was proposed and studied extensively by Hidalgo et. al in 2002 \cite{HKH02}. By construction it is a different class of fiber bundle model- there is no noise term and non-linearity in material response has been introduced through an exponent in the constitutive equation. This model is different from our simple noise-induced FBM. 
It has been observed that the strain-rate shows power-law relaxation  in the 
creep regime followed by a power-law acceleration up to complete rupture \cite{nhgs05}
and the waiting time distributions in  such creep models obey power laws \cite{BK08}.

Yoshioka et. al. \cite{YKI10,YKI12} discussed  thermally activated  failure in FBM introducing a Gaussian fluctuation in local force (stress) on individual fibers.  Potentially this model goes back to classical FBM if the fluctuation term is zero. But if the fluctuation is non-zero then the bundle can fail even when the  external stress is zero  which is confusing and  not real. 
In that sense our noise-induced failure scheme in FBM (introduced in 2003 in ref. \cite{Pradhan}) is more robust and some exact analytic results (failure time and avalanche distribution) have already been calculated through this scheme.

Identification of phase boundary is crucial for any dynamical system because 
a system usually changes its behavior as it moves from one phase to
another. As we can see in our model, there is no waiting time above the phase 
boundary (continuous rupture phase) and  waiting time appears below the phase
boundary (intermittent phase). One can also estimate the failure time of the 
system exactly \cite{Pradhan} in the continuous rupture phase. 
In case of fracturing in loaded rocks/materials, such study can help to identify
reliable precursors which can warn of an imminent breakdown. We notice,
in our model system, magnitude of waiting time reduces gradually towards the 
breakdown point which is reflected in the variation of $a$ in the functional 
form of the distribution. What is the exact form of this variation? Does it depend on the 
applied stress and noise level? Which one is the more sensitive parameter?
These questions must be answered to develop a prediction scheme based on 
available precursors prior to failure/breakdown.

This work is partially supported by Norwegian Research Council through grant 
no. 199970/S60 and 217413/E20. AKC thanks DST (India) for financial support under the Fast Track Scheme for Young Scientists Sanc. no. SR/FTP/PS-090/2010 (G).


\begin{thebibliography}{21}
\bibitem{CB97} B. K. Chakrabarti and L. G. Benguigui, {\it Statistical Physics 
of Fracture and Breakdown in Disordered Systems}, Oxford University Press 
(1997).
\bibitem{HR90} H. J. Herrmann and S. Roux, {\it Statistical Models for the 
Fracture of Disordered Media}, North-Holland, Amsterdam (1990). 
\bibitem{RMP} S. Pradhan, A. Hansen  and B. K. Chakrabarti, Rev. Mod. Phys. {\bf 82}, 499 (2010).
\bibitem{PC01} S. Pradhan and B. K. Chakrabarti, Phys. Rev. E {\bf 65}, 016113 (2001). 
\bibitem{PBC02}S. Pradhan, P. Bhattacharyya and B. K. Chakrabarti, Phys. Rev. E {\bf 66}, 016116 (2002).
\bibitem{BPC03} P. Bhattacharyya, S. Pradhan and B. K. Chakrabarti, Phys. Rev. E {\bf 67}, 046122 (2003).
\bibitem{Peirce} F. T. Peirce, J. Text. Ind. {\bf 17}, 355 (1926).
\bibitem{HH} P. C. Hemmer and A. Hansen, ASME J. Appl. Mech. {\bf 59}
 909 (1992).
\bibitem{PHH} S. Pradhan, A. Hansen and P. C. Hemmer, Phys. Rev. Lett. {\bf 95} 125501 (2005).
\bibitem{PH} S. Pradhan and P. C. Hemmer, Phys. Rev. E {\bf 75} 056112 (2007). 
\bibitem{Lawn} B. R. Lawn, {\it{Fracture of Brittle Solids}} (Cambridge University Press, Cambridge, 1993).
\bibitem{Coleman} B. D. Coleman, J. Appl. Phys. {\bf 29}, 968 (1958).
\bibitem{Ciliberto} R. Scorretti, A. Guarino and S. Ciliberto, 
Europhys. Lett. {\bf 55}, 626 (2001).
\bibitem{Roux} S. Roux, Phys. Rev. E {\bf 62}, 6164 (2000).
\bibitem{Pradhan} S. Pradhan and B.K. Chakrabarti, Phys. Rev. E {\bf 67}, 046124 (2003).
\bibitem{HKH02} R. C. Hidalgo, F. Kun and H. J. Herrmann Phys. Rev. E. {\bf 65}, 032502 (2002).
\bibitem{nhgs05} H. Nechad, A. Helmstetter, R. El Guerjouma and D. Sornette, 
Phys. Rev. Lett. {\bf 94}, 045501 (2005).

\bibitem{BK08} T. Baxevanis and T. Katsaounis, Eur. Phys. J. B. {\bf 61}, 153-157 (2008).
\bibitem{YKI10} N. Yoshioka, F. Kun and N. Ito Phys. Rev. E. {\bf 82}, 055102 (2010).
\bibitem{YKI12} N. Yoshioka, F. Kun and N. Ito Eur. Phys. Lett. {\bf 97}, 26006 (2012).
\bibitem{AE} A. Petri, G. Paparo, A. Vespignani, A. Alippi and M. Costantini,
 Phys. Rev. Lett. {\bf 73}, 3423 (1994).
\bibitem{Corral06} A. Corral, Phys. Rev. Lett. {\bf 97}, 178501 (2006).
\bibitem{Corral03} A. Corral, Phys. Rev. E. {\bf 68}, 035102 (2003).
\bibitem{Bak} P. Bak, K. Christensen, L. Danon and T. Scanlon, Phys. Rev. Lett. {\bf 88}, 178501 (2002).
\end{thebibliography}
\end{document}